# Efficient and Tunable Strip-to-Slot Fundamental Mode Coupling

Viphretuo Mere, Rakshitha Kallega, and Shankar Kumar Selvaraja

Photonics Research Laboratory, Centre for Nano
Science and Engineering, Indian Institute of science, Bangalore, 560012, India

shankarks@iisc.ac.in

**Abstract**

We present a novel photonic wire-to-slot waveguide coupler in SOI. The phase matching between a wire and slot mode is achieved using a mode transformer. The architecture consists of a balanced 50/50 power splitter and a tunable phase matched taper combiner forming a slot waveguide. We show a theoretical wire-to-slot coupling efficiency of 99 % is achievable and experimentally, we demonstrate a coupling efficiency of 99 % in the 1550 nm band. Based on the coupling scheme, we also show excitation of a slot mode in a slotted ring resonator and verified the excitation through the thermo-optic response of the rings. We show a nearly athermal behaviour of a PMMA filled slot ring with a thermo-optic response of 12.8 pm/° C compare to 43.5 pm/° C for an air clad slot waveguide.

## 1 Introduction

Integrated photonics, in particular, Silicon photonic is an excellent platform for sensing applications. It provides exemplary advantages of compactness, scalability, high sensitivity, and selectivity to realise an on-chip fluid and gas sensing platform [1, 2, 3, 4, 5]. Integrated photonic based sensors rely on light-matter interaction to detect the presence of an analyte. These interactions can result in absorption, scattering or emission of light; and when described as a function of wavelength, gives direct fingerprint of the analytes present [5]. Wire and Ridge waveguides were primarily used for surface sensing, however, in trace gas sensing, it is essential to have high light-matter interaction with the bulk of the analyte as well, for better sensitivity [6]. It has been shown that for larger light-matter interaction, slot waveguide is an ideal candidate [7]. Slot waveguide owing to its inherent ability to confine large fields (>40 percent) in low-index medium allows higher bulk interaction compared to wire/ridge waveguides. This has lead to exploitation of the slot-waveguide for non-linear, athermal device and sensing applications [8, 9, 10]. However, high-index contrast between Si and slot-gap results in higher propagation loss due to scattering along the sidewalls [11]. Nonetheless, it can be mitigated by post-fabrication treatments [12, 13]. Slot waveguide also suffers from coupling loss because of modal mismatch between wire and slot waveguide in direct butt coupling scheme. Recent reports shows various schemes to reduce coupling loss; asymmetric taper, inverted taper and Multi-Mode Interference coupler [14, 15, 16, 17]. However, these schemes require challenging device dimensions;< 50 nm, making it difficult to fabricate. Furthermore, device tolerance to fabrication imperfections restricts the applicability of the scheme[18].

In this paper, we propose a novel scheme where the coupling between a wire and slot waveguide is achieved using a split-and-combine scheme. The proposed scheme is the most versatile and robust wire-to-slot waveguide coupling reported with a coupling efficiency of -0.06 dB. We present a detailed design, tolerance analysis, experimental demonstration and verification of the slot mode excitation. We further demonstrate nearly athermal behaviour of the slot ring resonator by exploiting the high field confinement in the slot based on the low TOC 12 pm/°C of a slotted ring resonator with polymethyl-methacrylate (PMMA) a negative thermo-optic material as an upper clad.

## 2 Design and simulations

A slot waveguide consists of two strip/wire high-index rail separated by a low-index gap [7]. At the boundary between the low- and high-index medium, the normal component of the Transverse Electric field (TE) redistributes to maintain continuity, which causes higher field confinement in the low-index material. The field enhancement inside the slot is directly proportional to the index contrast between the rail and the gap. Figure 1 shows the E-field distribution and effective index of a conventional wire waveguide and a slot waveguide in SOI. Since the mode field and the propagation constant of the wire and slot modes are different, coupling light between these two is a challenge. Figure 2(a) shows the proposed wire-to-slot coupling scheme.

The scheme consist of a wire waveguide that is split into two single mode waveguides with a Y-splitter. The split phase matched waveguides are then brought close together to form a slot waveguide. Figure 2(b) shows the 3D FDTD simulation of the proposed coupler.

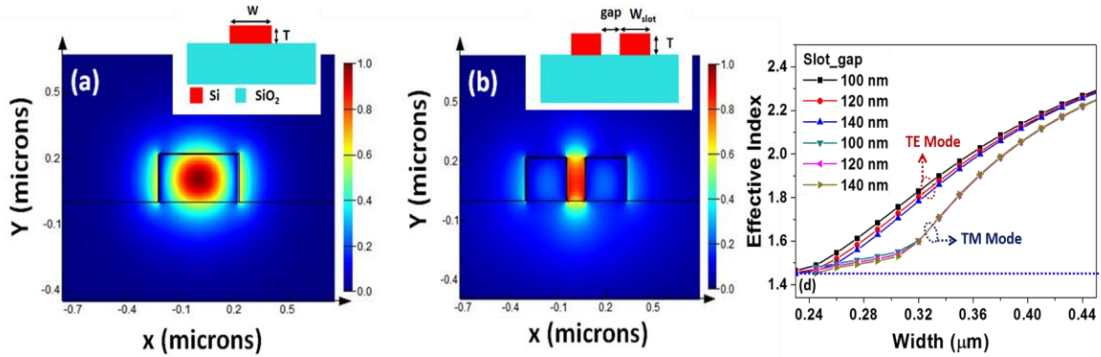

Figure 1: Electric field distribution profile of the two waveguide configurations, (a) Wire waveguide (W = 450 nm, T = 220 nm), (b) Slot waveguide (W = 260 nm, gap = 120 nm, T = 220 nm), and (c) Effective index of slot waveguide (air clad) as a function of slot width (Wslot) and gap.

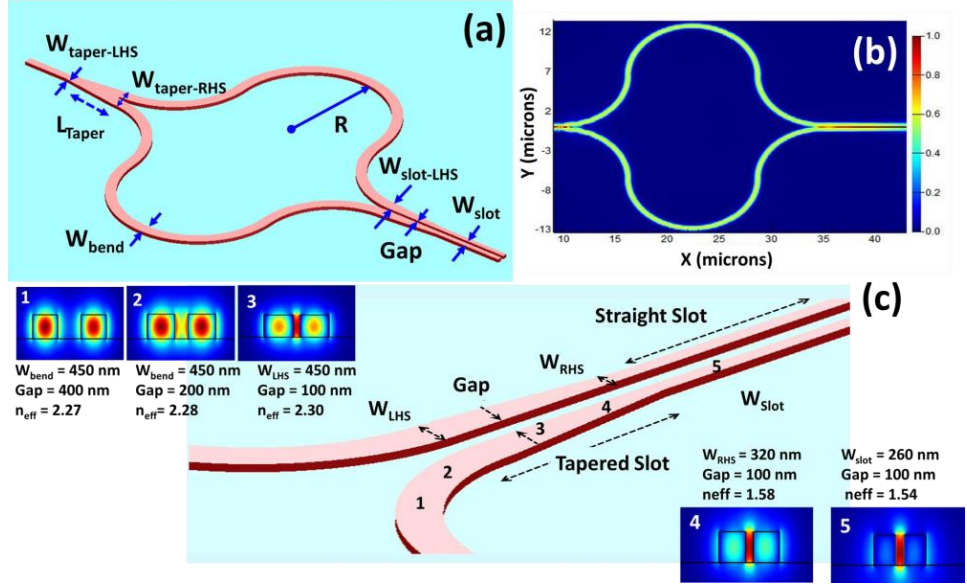

Figure 2: (a) Schematic of the proposed wire-to-slot coupler, (b) Simulated E-field (V/m) distribution (3D FDTD) and excitation 100 nm gap slot waveguide mode, and (c) Effective index and electric field evolution at different cross-section points along the length of the coupler.

Figure 2(c) (inset 1-5) illustrates the effective index ($n_{eff}$) and E-field evolution at five different cross sections along the length of the coupler. As the two wire waveguides were brought closer to form a slot, the system behaves like a single waveguide where the mode is distributed between the two waveguides resulting in higher $n_{eff}$ Figure 2(c) (inset 1,2,3). As the waveguide width is linearly tapered from 450 nm to 260 nm, the $n_{eff}$ decreases resulting in phase matching with the slot mode Figure 2(c) (inset 4, 5). An optimised coupler has a simulated coupling efficiency of 99.45, 96.01 and 95.54 % for a slot gap of 100, 120 and 140 nm respectively. The Coupling Efficiency (CE) is calculated as the total power at the output (Pout) to the total power in the two bend waveguides ($P_{arm1}$ + $P_{arm2}$), given as, $CE = P_{out}/(P_{arm1} + P_{arm2})$. It can be observed that coupling efficiency as high as 99% can be achieved with the proposed scheme. For the efficiency calculation, ideal power splitting scenario is considered. Nevertheless, fabrication and thickness non-uniformity can lead to reduction in the coupling efficiency due to phase mismatch in the arms. The robustness and tunable nature is presented in the following section.

## 2.1 Sensitivity and tolerance

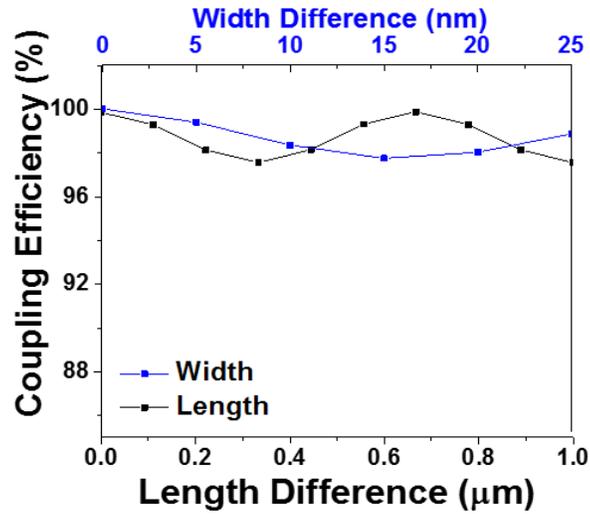

Figure 3: Effect of width and length variation of the phase matched arms on the coupling efficiency.

Unlike any reported couplers, the proposed configuration is tunable. Any fabrication non-uniformity in the phase matched waveguides could result in a reduction in the coupling efficiency. However, one can use thermo-optic tuning in the split waveguide section to compensate any phase mismatch. Figure 3 shows the effect of coupling efficiency when the length of one of the arm is changed while rest of the parameters were kept optimal and constant. The change in the length or width of the split waveguides creates a phase imbalance and result in non-optimal excitation or coupling of the slot mode. For a length difference of 1 m between the arms, we observe only a 3% reduction in the coupling efficiency. Similarly, the effect of waveguide width variation also results in 3% reduction in coupling efficiency for a linewidth variation of 25 nm, which is more than 5% variation.

# 3 Experimental and Measurements

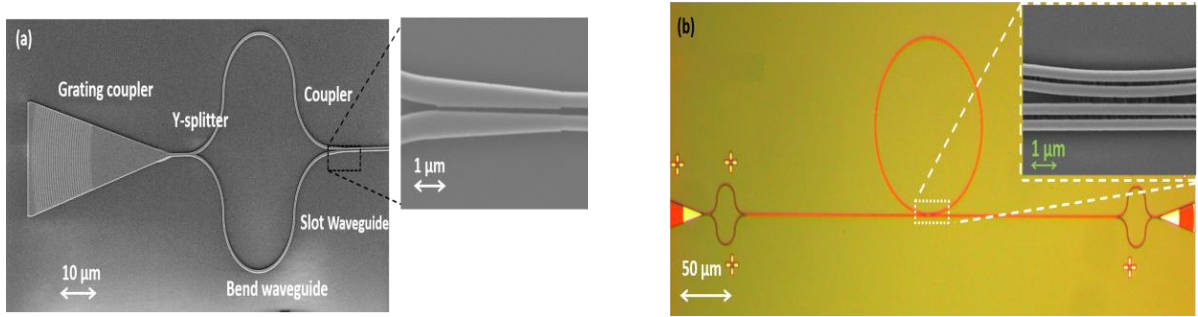

Figure 4: (a) SEM image of a fabricated wire-to-slot coupler with grating coupler and (b) Optical image of a slot ring resonator (Wslot/gap = 260/140 nm); inset: SEM image of the coupling region between the slotted bus waveguide and ring resonator.

The device is fabricated in an SOI wafer of 220 nm device layer on 2 m buried oxide. To demonstrate the coupling scheme, a test structure consisting of a wire-to-slot coupler along with a slotted ring resonator was designed and fabricated. The light is coupled in and out of the chip using shallow etched grating coupler. The pattering is done using electron beam lithography and Inductively Coupled Plasma-Reactive Ion Etch (ICP-RIE) process. The waveguides and gratings were patterned using ma-N-2401 and PMMA-A4 920 resist followed by 220 nm and 70 nm Si etch using ICP-RIE process. Both the etch process is done using $C_4F_8$ and $SF_6$ gas chemistry. Figure 4(a) and 4(b) shows the fabricated wire-to-slot waveguide coupler and microscope image of a 60 m diameter slotted ring resonator in SOI.

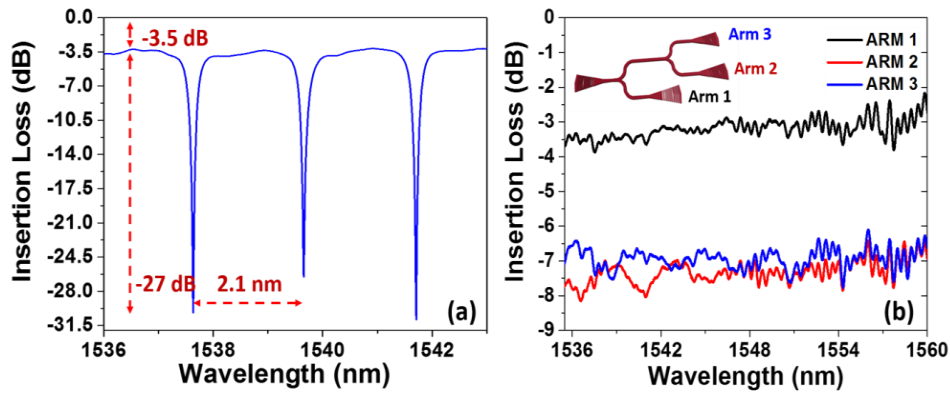

Figure 5: (a) Spectral response of a PMMA filled slotted ring resonator; (b) Insertion loss measured at three different Y-splitter arms.

The fabricated slot waveguide devices were characterised using a tunable laser and a power in the 1550 nm band. Figure 5(a) shows the normalised transmission spectrum of the fabricated slotted ring where the slot waveguide is excited by the proposed wire-to-slot coupler. Normalisation removes the loss due to fibre-chip couplers. We observed an insertion loss of ~-3.5 dB with a maximum extinction ratio of -27 dB. We measure a group index of ~3.17 around 1550 nm which agrees well with the simulated fundamental slot mode group index of 3.12. The group index confirms the excitation of the propagating slot mode by the proposed coupler. We attribute the insertion loss to the propagation loss in the slot waveguide and Y-splitter. Figure 5(b) shows the insertion loss of the Y-splitter extracted from the Y-splitter chain shown as an inset in Figure 5(b). An insertion loss of 0.5 dB is measured per splitter/combiner. In the test circuit, we use a splitter and a combiner resulting in an insertion loss of -1 dB. On the other hand, the propagation loss of slot waveguide is determined by using cutback method. A propagation loss of -2.44 dB was measured for a 300 m long slot waveguide that was used in the test circuit. Therefore, by normalizing the ring response with respect to the insertion loss of Y-splitter (1 dB) and slot waveguide propagation loss (2.44 dB), we can conclude that the proposed strip-to-slot coupler design has an insertion loss of 0.06 dB or a coupling efficiency of 99%.

## 4 Temperature dependence of Slot waveguides

In this section, we present an additional verification of the slot mode excitation using the thermo-optic property of the slot waveguide. It is known that Si has a high TOC ($(1.86 \times 10^{-4}/°C)$), however, by using a negative TOC cladding material one can realise am athermal circuit. It has been shown, athermal circuits can be built by infusing a negative thermo-optic coefficient (TOC) material in the slot waveguide [8, 19, 20]. Athermal circuits are essential for circuits to operate in dynamically varying thermal scenarios. For a given waveguide system, the equivalent thermo-optic coefficient (TOC) of a waveguide system is a superposition of the TOC of the individual materials, which is given by [8],

$$\frac{dn_{eff}}{dT} = \sum_i \frac{d(\Gamma_i n_i)}{dT} \approx \sum_i \Gamma_i \frac{d(n_i)}{dT} = \Gamma_{core}\frac{d(n_{core})}{dT} + \Gamma_{bot-clad}\frac{d(n_{bot-clad})}{dT} + \Gamma_{up-clad}\frac{d(n_{up-clad})}{dT}$$

(1)

Where, $\frac{dn_i}{dT}$ is the TOC of the i-th material. The TOC of the core, upperclad, bottom-clad are represented as $\frac{d(n_{core})}{dT}, \frac{d(n_{bot-clad})}{dT}, \frac{d(n_{up-clad})}{dT}$, respectively. $\Gamma_i$ is the confinement factor of the individual materials. Equation (1) indicates that the overall TOC is highly dependent on the confinement factor. Figure 6(a) shows the $n_{eff}$ at various temperature of a PMMA filled slot waveguide. Figure 6(b) shows the confine factor as a function of slot waveguide gap. As the gap decreases the light confined in the slot increases. With a negative TOC material as top clad, simulated thermo-optic response shows an athermal operation at a slot gap of 120 nm. For gaps larger than 120 nm, we observe an under compensation while for smaller gaps we observe an over compensation. The compensation depends on the magnitude of the negative TOC of the cladding material, in this example PMMA with a TOC of $-1 \times 10^{-4}/°C$ is used for the calculations. Taking ring resonator as an example, we investigate the shift in the resonant wavelength with temperature change. The thermal response of the resonant wavelength ($\lambda_r$) is expressed as follows [19].

$$\frac{d\lambda_r}{dT} = (n_{eff}\alpha_{sub} + \frac{dn_{eff}}{dT}) = \frac{dn_{eff}}{dT}\frac{\lambda_r}{n_g} \qquad (2)$$

Where, $\lambda_r$ is the resonant wavelength, $n_g$ is the group index. $\alpha_{sub}$ is the thermal expansion coefficient of the substrate. Since the magnitude of $\alpha_{sub}$ ($10^{-7}/°C$) is small, the influence of thermal expansion of the substrate ($SiO_2$) on the resonance shift is minimal. The $\frac{d\lambda_r}{dT}$ for the fundamental TE mode is calculated by substituting $\frac{dn_{eff}}{dT}$ into eqn. (2); where, the values of $\frac{dn_{eff}}{dT}$ is calculated using a mode-solver. Figure 6(b) shows the athermal operation $\frac{d\lambda_r}{dT}$ is zero at slot width and gap of 240 nm and 120 nm respectively.

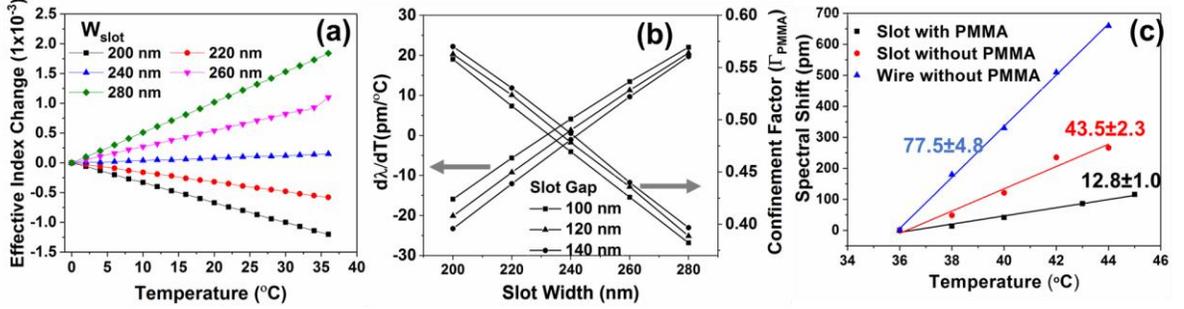

Figure 6: (a) Simulated slot-gap influence on the TO effect of the slot waveguide with a slot gap of 120 nm, (b) Evanescent field fraction and temperature dependency of the resonance of a PMMA filled slotted ring resonator and (c) Experimentally TO response of a slot and wire waveguide ring resonator.

The slotted ring resonator fabricated earlier was covered with PMMA as upper clad and the resonance characteristics is measured at different chip temperatures. In this case, the chip was loaded on a temperature Table 1: Thermal response of slot and wire waveguide based ring resonator.

| Waveguide configuration | Coupling Efficiency (%) | |
|---|---|---|
| | Experimental | Simulation |
| Slot with PMMA | 12.8 ± 1.0 | 9.417 |
| Slot Without PMMA | 43.5 ± 2.3 | 44.46 |
| Wire Withour PMMA | 77.5 ± 4.8 | 77.74 |

controlled stage. Figure 6(c) shows the thermo-optic reponse of a PMMA filled ring along with a wire waveguide and slotted ring without PMMA for comparison. It can be clearly observed that PMMA filled slotted rings shows the minimal response to temperature change, showing a nearly athermal behaviour of ~12.8 pm/°C. Table 1 compares the closely matching simulated and measured thermo-optic response.

## 5 Conclusion

We have presented a novel wire-to-slot mode excitation scheme using a splitter and a tapered slot waveguide coupler. A coupling efficiency of 99% from a wire waveguide to slot waveguide is achieved at 140 nm slot gap. Experimentally we have shown slot mode excitation and verified it using thermooptic behaviour of a slotted ring resonator. A low thermal dependence of slot waveguide ring resonator of 12 pm/°C is achieved, indicating high field concentration in the slot region.

## References

[1]  E. Ryckeboer, R. Bockstaele, M. Vanslembrouck, and R. Baets, "Glucose sensing by waveguide-based absorption spectroscopy on a silicon chip," **5**(5), 1636–48, (2014).

[2]  Y. Huang, S. K. Kalyoncu, Q. Zhao, R. Torun, and O. Boyraz, "Siliconon-sapphire waveguides design for mid-IR evanescent field absorption gas sensors" **313**, 186–194, (2014).

[3]  A. Dhakal, P. Wuytens, A. Raza, N. Le Thomas, and R. Baets, "Silicon nitride background in nanophotonic waveguide enhanced Raman spectroscopy," Materials **10**(2), (2017).

[4]  D. M. Kita, H. Lin, A. Agarwal, K. Richardson, I. Luzinov, T. Gu, and J. Hu, "On-Chip Infrared Spectroscopic Sensing: Redefining the Benefits of Scaling," **11**(12), 1385–1391, (2003).

[5]  A. Z. Subramanian, E. Ryckeboer, A. Dhakal, F. Peyskens, A. Malik, B. Kuyken, H. Zhao, S. Pathak, A. Ruocco, A. De Groote, P. Wuytens, D. Martens, F. Leo, W. Xie, U. D. Dave, M. Muneeb, P. Van Dorpe, J. Van Campenhout, W. Bogaerts, P. Bienstman, N. Le Thomas, D. Van Thourhout, Z. Hens, G. Roelkens, and R. Baets, "Silicon and silicon nitride photonic circuits for spectroscopic sensing on-a-chip," **3**(5), B47–B59, (2015).

[6]  T. Claes, J. G. Molera, K. De Vos, E. Schacht, R. Baets, and P. Bienstman, "Label-free biosensing with a slot-waveguide-based ring resonator in silicon on insulator," IEEE Photonics J. **1**(3), 197–204, (2009).

[7]  V. R. Almeida, Q. Xu, C. A. Barrios, and M. Lipson, "Guiding and confining light in void nanostructure," **29**(11), 1209–1211, (2004).

[8]  Feng, K. Shang, J. T. Bovington, R. Wu, B. Guan, K.-T. Cheng, J. E. Bowers, and S. J. Ben Yoo, "Athermal silicon ring resonators clad with titanium dioxide for 13 m wavelength operation," **23**(20), 25653– 25660, (2015).

[9]  L. Zhou, K. Okamoto, and S. J. B. Yoo, "Athermalizing and trimming of slotted silicon microring resonators with UV-sensitive PMMA uppercladding," IEEE Photon. Technol. Lett. **21**(17), 1175–1177, (2009).

[10] C. Koos, P. Vorreau, T. Vallaitis, P. Dumon, W. Bogaerts, R. Baets, B. Esembeson, I. Biaggio, T. Michinobu, F. Diederich and W. Freude,"All-optical high-speed signal processing with silicon organic hybrid slot waveguides," Nature photonics **3**(4), 216–219 (2009).

[11] T. Baehr-Jones, M. Hochberg, C. Walker, and A. Scherer, "High-Q optical resonators in silicon-on-insulator-based slot waveguides," Appl. Phys. Lett. **86**(8), 1–3, (2005).

[12] T. Alasaarela, D. Korn, L. Alloatti, A. Säynätjoki, A. Tervonen, R. Palmer, J. Leuthold, W. Freude, and S. Honkanen, "Reduced propagation loss in silicon strip and slot waveguides coated by atomic layer deposition," **19**(12), 11529–11538, (2011).

[13] D. K. Sparacin, S. J. Spector, and L. C. Kimerling, "Silicon waveguide sidewall smoothing by wet chemical oxidation," J. Light. Technol. **23**(8), 2455–2461, (2005).

[14] N. N. Feng, R. Sun, L. C. Kimerling, and J. Michel, "Lossless strip-toslot waveguide transformer," **32**(10), 1250–1252, (2007).

[15] Z. Wang, N. Zhu, Y. Tang, L. Wosinski, D. Dai, and S. He, "Ultracompact low-loss coupler between strip and slot waveguides," **34**(10), 1498–1500, (2009).

[16] A. Säynätjoki, L. Karvonen, X. Tu, T. Y. Liow, A. Tervonen, G. Q. Lo, and S. Honkanen, " Compact and efficient couplers for silicon slot waveguides," In SPIE OPTO, pp. 82660J–82660J. International Society for Optics and Photonics, (2012).

[17] Q. Deng, L. Liu, X. Li, and Z. Zhou, "Strip-slot waveguide mode converter based on symmetric multimode interference," **39**(19), 5665–5668, (2014).

[18] V. M. N. Passaro and M. la Notte, "Optimizing SOI slot waveguide fabrication tolerances and strip-slot coupling for very efficient optical sensing," Sensors **12**(3), 2436–2455, (2012).

[19] J. Teng, P. Dumon, W. Bogaerts, H. Zhang, X. Jian, X. Han, M. Zhao, G. Morthier, and R. Baets, "Athermal Silicon-on-insulator ring resonators by overlaying a polymer cladding on narrowedwaveguides," **17**(17), 14627–14633, (2009).

[20] A. Säynätjoki, T. Alasaarela, A. Khanna, L. Karvonen, P. Stenberg, M. Kuittinen, A. Tervonen, and S. Honkanen, "Angled sidewalls in silicon slot waveguides: conformal filling and mode properties." **17**(27), 21066–21076, (2009).